# An Upper Limit of AC Huffman Code Length in JPEG Compression

Kenichi Horie*

R&D Division, Olympus Imaging Corporation, 2951 Ishikawa, Hachioji, 192-8507 Tokyo, Japan

E-mail: kenichi_horie@ot.olympus.co.jp

**Abstract.** A strategy for computing upper code-length limits of AC Huffman codes for an 8x8 block in JPEG Baseline coding is developed. The method is based on a geometric interpretation of the DCT, and the calculated limits are as close as 14% to the maximum code-lengths. The proposed strategy can be adapted to other transform coding methods, e.g., MPEG 2 and 4 video compressions, to calculate close upper code length limits for the respective processing blocks.

## I. INTRODUCTION

JPEG Baseline coding [1] is by far the most widely implemented method for coding still images. Statistical behavior of the DCT coefficients and the correlations between the runs of zeros and the sizes of coefficients have been studied, which resulted in the JPEG 2-dimensional Huffman coding [2], [3]. Furthermore, the distribution of the DCT coefficients has also been investigated in detail [3]-[5]. While these statistical behaviors of JPEG are well known, little is known about the mathematical bounds on the code-lengths. Of particular interest are the AC Huffman codes, which exhibit large variations in code-length and thereby determine the most important part of the total code-length of a JPEG file.

The mathematical study of upper limits of AC Huffman code-length is not only of academic interest, but stems from needs in consumer electronics. Most digital cameras employ the JPEG Baseline sequential coding, and the knowledge of a close upper code-length limits for an 8x8 block (and the JPEG image file) is important in assigning economic buffer memory spaces at various stages of image encoding in a digital camera.

Furthermore, a digital camera usually displays the number of images fitting into the rest of the memory space, and to ensure that this number of JPEG images can indeed be stored, the quantization tables for each JPEG image are optimized to keep the file size close to or below a targeted value. However, such a rate control is achieved by iterating the quantization, the Huffman encoding, and the bit counting of coded data length [6] or by other, sophisticated file size prediction schemes [7], both of which require extensive processing. Alternatively, rate control can be avoided by calculating the number of images based on an average JPEG file size and its standard variation [8]. However, such a statistical estimation method is not reliable, especially in a repetitive shooting mode wherein successive images are often statistically correlated with each other. With the knowledge of a close upper limit of the JPEG file size, a rate control scheme is not necessary anymore, or at least its use greatly reduced. Additionally, the reliability of statistical methods for calculating the number of images can be significantly improved by taking the upper limit into account.

A JPEG file may be formatted according to JFIF [9] or Exif [10] and consists of AC Huffman

codes, the DC Huffman codes, the Huffman and the quantization tables, and other portions including a header. Close upper code-length limits or even the maximum code-lengths of those portions except the AC Huffman codes can be easily calculated. On the other hand, it is extremely difficult to calculate a close limit of the AC Huffman code-length. One reason for this difficulty is that the AC Huffman codes are assigned to the runlength-size combination, which extends over several DCT coefficients rather than to individual coefficients. A further, subtle reason is that little is known about the space of the DCT coefficients and the functional behavior of the code-length in this space.

In this work we clarify the geometrical aspect of the DCT coefficients to derive a computational method for calculating a close upper limit on the AC Huffman code-length for an 8x8 block. The calculation method is derived in dependence on the scaling of the example quantization tables in annex K of the JPEG specification [1].

Although only the JPEG case is described in this work, the general principles of our method are applicable to other transform coding methods. In particular, these approaches can be adapted to calculate close upper limits of the VLC code-lengths in MPEG 2 and MPEG 4. Some of the proposed ideas might be applicable even for H.264/AVC, VC-1, and JPEG XR video/still image compressions.

## II. JPEG BASELINE CODING AND THE GEOMETRY OF THE DCT COEFFICIENTS

We introduce terms and definitions used in this work and thereby briefly review some aspects of the JPEG Baseline encoding process [1]-[3]. We also add mathematical comments and observations which are necessary in introducing and developing our calculation strategy.

In the JPEG Baseline coding a (color) image consists of, for example, one luminance (Y) and two chrominance (Cb, Cr) matrices of 8-bit pixel data. Each of these matrices is sectioned into 8x8 blocks. In each pixel block, its pixel values are shifted from unsigned integers in the range $[0, 2^8 - 1]$ to signed integers in the range $[-2^7, 2^7 - 1]$. Let $f_{xy}$ ($x, y = 0,...,7$) denote the shifted pixel values of an 8x8 pixel block, which can be viewed as a vector in a 64-dimensional space. The set of all possible vectors furnishes a cube $I = [-2^7, 2^7-1]^{64}$, which is off-centered from the coordinate origin. This cube in turn is contained in a ball $B_0$ around the coordinate origin with radius $R = 2^7 \times \sqrt{64} = 2^{10}$.

Let $K_{xu} = \frac{1}{2} C(u) \cos \frac{(2x+1)u\pi}{16}$ be the Discrete Cosine Transform (DCT) matrix, where $x, u = 0, \ldots, 7$, and $C(u) = 1/\sqrt{2}$ for $u = 0$ and $C(u) = 1$ for $u = 1, \ldots, 7$. The DCT coefficients of the pixel block are given by the expression $F_{uv} = \sum_{x,y=0}^{x,y=7} K_{xu} f_{xy} K_{yv}$. In matrix notation this reads $F = K^T f K$, where the superscript $T$ denotes the transpose of a matrix. Among the DCT coefficients $F_{uv}$, the upper-left coefficient $F_{00}$ is 8 times the average value of all pixel data within the 8x8 pixel block and is called the DC coefficient. The remaining 63 coefficients are called AC coefficients. We call $F_{uv}$ or $F$ a *(DCT) configuration*, and $f_{xy}$ or $f$ simply a *vector*.

Although the DCT is introduced in literature as a mapping from the image domain into the spatial frequency domain, we can interpret the DCT simply as an orthogonal mapping acting upon vectors within the 64-dimensional space. To see this, we use the fact that the DCT matrix is an orthogonal matrix, i.e. $KK^T = K^T K = I$. This can be derived naturally by noting that these column vectors are eigenvectors of a real symmetric second difference matrix, which automatically ensures that the column vectors are orthogonal [11]. In order to verify the orthogonal feature of the DCT as a 64-dimensional transformation, we have to show that the natural inner product in the 64-dimensional space is preserved under the DCT. The inner product between two general 64-dimensional vectors $v$ and $w$ is given by the trace $Tr(v \cdot w^T)$, wherein the vectors are assumed to take the form of an 8x8 matrix. Let $V = K^T v K$ and $W = K^T w K$ be the image vectors under the DCT. Using the orthogonality of the DCT matrix and the invariance of the trace under cyclic permutation of the matrices, it follows $Tr(VW^T) = Tr(K^T v K K^T w^T K) = Tr(vw^T)$.

Since the DCT is orthogonal, it is a 64-dimensional rotation which rotates the cube $I$ to another cube $J$, but leaves the ball $B_0$ unchanged. We call $J$ the *configuration space*, which by definition contains all possible DCT configurations and is a subset of $B_0$. Since the DCT configuration $F_{uv}$ lies in the ball $B_0$, the coefficients satisfy the relation $\sum_{u,v} (F_{uv})^2 \leq 2^{20}$. In particular, all coefficients must lie in the range of $[-2^{10}, 2^{10}]$. As we will be focusing on the AC coefficients in the subsequent sec-tions, we need further characterizations of the AC coefficients with respect to the ball. We have:

Proposition 1. *Regardless of the DC coefficient value, the AC coefficients fulfill the following strict inequality:*

$$\sum_{(u,v) \neq (0,0)} (F_{uv})^2 < 2^{20} \quad \text{(AC Ball condition)}. \tag{1}$$

*Proof.* If this is wrong, a DCT configuration $F$ would exist with the property $\sum_{(u,v) \neq (0,0)} (F_{uv})^2 = 2^{20}$. Then the DC coefficient $F_{00}$ must be necessarily zero, and $F$ lies on the sphere surface $S$ (with radius $R$) of the ball $B_0$. Since the DCT is a rotation, the inverse image vector $f$ must also lie on the sphere $S$. However, due to the off-centered nature of the cube $I$, the sum of the squared amplitudes of the pixel data $f_{xy}$ cannot become the square of the radius $R$ unless all pixel data are equal to $-2^7$. This implies that the image vector $f$ is flat, with no spatial frequency, so that under the DCT, all AC coefficients would become zero, and the DC coefficient would have the value $-2^{10}$, contradicting the assumption and proving the assertion.

The inequality (1) means geometrically that DCT configurations cannot lie on the 62-dimensional equator $E$ of the sphere $S$, which is defined by the set of the two equations $F_{00} = 0$ and $\sum_{(u,v) \neq (0,0)} (F_{uv})^2 = 2^{20}$. Thus the DCT configurations lie in the "half-open" ball $B = B_0 \setminus E$, in which this equator has been taken away from the ball $B_0$. Henceforth we call this half-open ball $B$ simply the *Ball*.

Expressing the condition that a configuration $F_{uv}$ lies in the configuration space $J$ is not an easy task because the rotated cube $J$ is not symmetric with respect to the coordinate axes. Instead, it is easier to apply the inverse DCT to $F_{uv}$ and to express that the inverse vector lies in the cube $I$. More precisely, the inverse vector must lie on the integer coordinate lattice within the cube $I$:

$$-2^7 \leq \sum_{u,v=0}^{u,v=7} K_{xu} F_{uv} K_{yv} \leq 2^7 \quad \text{for all} \quad x, y = 0,...,7 \quad \text{(Cube Condition)} \tag{2}$$

$$\sum_{u,v=0}^{u,v=7} K_{xu} F_{uv} K_{yv} \quad \text{is integer for all} \quad x, y = 0,...,7 \quad \text{(Integer Condition)} \tag{3}$$

The DCT coefficients are quantized with an integer-valued quantization table $Q_{uv}$ ($\geq 1$). The quantized DCT coefficients can then be calculated as $D_{uv} = \text{INT}(F_{uv}/Q_{uv})$, where INT denotes the integer portion (one may also use the rounding function instead). Due to the AC Ball condition (1), the AC coefficients cannot become $-2^{10}$ nor $2^{10}$, and the quantized AC coefficient values can be constrained to lie in the interval $[-2^{10}+1, 2^{10}-1]$.

| 16 | 11 | 10 | 16 | 24 | 40  | 51  | 61  |
|----|----|----|----|----|-----|-----|-----|
| 12 | 12 | 14 | 19 | 26 | 58  | 60  | 55  |
| 14 | 13 | 16 | 24 | 40 | 57  | 69  | 56  |
| 14 | 17 | 22 | 29 | 51 | 87  | 80  | 62  |
| 18 | 22 | 37 | 56 | 68 | 109 | 103 | 77  |
| 24 | 35 | 55 | 64 | 81 | 104 | 113 | 92  |
| 49 | 64 | 78 | 87 | 103| 121 | 120 | 101 |
| 72 | 92 | 95 | 98 | 112| 100 | 103 | 99  |

Table K.1 – Luminance quantization table

| 17 | 18 | 24 | 47 | 99 | 99 | 99 | 99 |
|----|----|----|----|----|----|----|----|
| 18 | 21 | 26 | 66 | 99 | 99 | 99 | 99 |
| 24 | 26 | 56 | 99 | 99 | 99 | 99 | 99 |
| 47 | 66 | 99 | 99 | 99 | 99 | 99 | 99 |
| 99 | 99 | 99 | 99 | 99 | 99 | 99 | 99 |
| 99 | 99 | 99 | 99 | 99 | 99 | 99 | 99 |
| 99 | 99 | 99 | 99 | 99 | 99 | 99 | 99 |
| 99 | 99 | 99 | 99 | 99 | 99 | 99 | 99 |

Table K.2 – Chrominance quantization table

The JPEG Baseline coding allows control over the file size by varying the quantization table. A frequently employed method consists in multiplying a scale factor $SF$ to the example quantization tables K.1 and K.2 of the JPEG specification to obtain scaled quantization tables:

$$Q_{uv} = \max(\text{INT}(SF \times Q_{0uv}), 1). \tag{4}$$

The smaller the scale factor, the finer the quantization and the larger the JPEG file size. When $SF = 1/64$, all quantization factors become one, whereas when $SF = 1$, we have the original example quantization tables K.1 and K.2, whose largest quantization factor is 121. In this work we consider scale factors between 1 and $1/64$.

After the quantization the quantized AC coefficients are ordered in a zigzag scan order and are represented by runlength, size, and amplitude. Let $D(k)$ denote the quantized DCT coefficients in the zigzag scan order of the JPEG specification, where $k = 0,...,63$ is the index in the zigzag ordering. The DC coefficient is $D(0)$. In what follows, we also use the notations $F(k) := F_{uv}$ and $Q(k) := Q_{uv}$ for the unquantized coefficients and the quantization values, respectively, in the zigzag ordering. For each nonzero quantized AC coefficient $D(k)$, $k = 1,...,63$, the runlength (consecutive number, possibly zero) of zero-valued AC coefficients which precede it in the zigzag sequence is determined. Each such runlength/nonzero-coefficient combination is represented by a duplet of integer numbers $(r,s)$ (symbol) together with the amplitude $a$ of the nonzero-coefficient. Here $r$ is the runlength $0$ to $15$, and the size $s$ is the number of bits used to encode the amplitude in the signed-integer encoding method of JPEG, so that $2^{s-1} \leq |a| < 2^s$ and $s = 1,...,10$. For zero-runs greater than $15$, the extension duplet $(15,0)$ is interpreted as a succession of $16$ zeros. Up to three consecutive $(15,0)$ extensions may precede a duplet $(r,s)$ with the total runlength $16 \times 3 + r$. If the last run of zeros includes the last (63rd) AC coefficient, a special duplet $(0,0)$ is assigned to this run of zeros to indicate the EOB (end of block), which terminates the 8x8 block. If the last AC coefficient is not zero, the EOB symbol is not assigned. In this work we re-define the size $s$ to be the smallest integer satisfying $|a| < 2^s$. This definition is the same as the usual definition for $s = 1,...,10$. Additionally, it introduces the zero size $s = 0$, which implies that the quantized coefficient $D(k)$ is in fact zero.

In the JPEG encoding process, the sequence $D(k)$ of quantized AC coefficients are represented by a sequence of pairs $(r,s)a$ of symbols $(r,s)$ and amplitudes $a$ in ascending order of the zigzag scan: $(r_1,s_1)a_1,(r_2,s_2)a_2,(r_3,s_3)a_3,...$. Huffman codes $Huff(r,s)$ of length up to 16 bits are assigned to each symbol $(r,s)$, whereas each amplitude $a$ is encoded with $s$ bits. The EOB is 4 (2) bits long for luminance (chrominance) AC coefficients. Throughout this work we use the typical Huffman tables in annex K of the JPEG specification [1].

Let $len$ denote the *code-length in bits* of a (sub-)sequence of AC coefficients, represented as pairs of symbols $(r,s)$ and amplitudes $a$ in the ascending zigzag scan order, or of Huffman codes $Huff(r,s)$ themselves. Since each symbol $(r,s)$ already determines the code-length of the pair $(r,s)a$, we define $len(r,s)$ to be the code-length of such a pair. For a coefficient $D(k) = a$ with size $s > 0$ and preceded by $r$ zeros, the relation $len(0,0,...,0,D(k)) = len(r,s) = len(Huff(r,s)) + s$ holds. For example, the zero runlength sequence $0,0,0,0,-1$ of chrominance AC coefficients results in $len(0,0,0,0,-1) = len(4,1) = len(Huff(4,1)) + 1 = len("111010") + 1 = 7$, see Table I. Since the code-length depends on the amplitudes only through the respective sizes, we also use a simplified notation $len(s_1, s_2, ..., s_l)$ for a sequence of AC coefficients having coefficient sizes $s_1, s_2, ..., s_l$ (some of which might be zero). Furthermore, $len(F)$ denotes the (luminance or chrominance) AC code-length of a configuration $F$.

The typical Huffman tables are structured such that $len(r,s) > len(r',s')$ for $r \geq r' \geq 0$ and for $s > s' > 0$, see Tables I and II below. Therefore, the AC code-length $len(F)$ of a configuration $F$ increases when the quantized AC amplitudes increase and/or the runlengths increase. In particular, one might expect that, at least in general, the code-length increases when the *unquantized* AC amplitudes increase, so that a configuration with the maximum code-length would lie on the boundary of the configuration space $J$. More precisely, we have the following fact:

Proposition 2. *For any maximum code-length configuration, there exists a configuration on the boundary having the same code-length. Furthermore, a maximum code-length configuration lies always on the boundary of $J$ or at most in such a distance away from it where the code-length does not change.*

*Proof.* Assume that $F$ is a maximum code-length configuration. Certainly then, not all AC coefficients of $F$ can become zero after quantization, so that at least one coefficient $F_{uv}$ remains non-zero after quantization. If $F$ is not already on the boundary of $J$, we can gradually enlarge the amplitude of $F_{uv}$ until we hit the boundary and thereby obtain a configuration $F'$ on the boundary. During this process the code-length $len(F)$ cannot become smaller due to the above mentioned monotony structure of the code-lengths $len(r,s)$, but also cannot become larger, since $F$ is assumed to be a maximum code-length configuration. Thus $F$ has the same code-length as $F'$, which proves the assertion.

Since the arguments do not depend on the specific geometry of $J$, the same statements are true if we consider configurations in the Ball $B$.

Having discussed the coding of AC coefficients, we briefly review the DC coefficients. The DC coefficients are encoded separately from the AC coefficients and take values in the range $[-2^{10}, 2^{10}]$. The quantized DC coefficient is encoded as the difference from the DC term of the previous block in the raster-scan order. The difference is represented by the amplitude $a$ and its size $s = 1,...,11$, $2^{s-1} \leq |a| < 2^s$. The size is Huffman coded, and the amplitude is encoded with $s$ bits. The maximum amplitude of the difference between two quantized DC coefficients is given by $\text{INT}(2 \times 2^{10} / Q_{00})$, where $Q_{00}$ is the DC quantization value.

## III. GENERAL STRATEGY

The AC code-length of a DCT configuration can certainly not exceed 63 times the maximum code-length in the Huffman table plus the length of the EOB code. The maximum length of Huffman codes is 16 bits, the maximum size is 10, and the EOB is 4 or 2 bits long. Thus a crude upper bound in bits for AC code-length is given by

$$l_0 = 63 \times (16 + 10) + 4 = 1642. \tag{5}$$

Besides being obviously too high and therefore useless, this number was derived with no reference to the quantization table. Contrary to the case of the DC coefficients, the AC coefficients are encoded in context with each other in combination of runlength and size, and the underlying geometry of the configuration space is rather complicated. These make it very difficult to find a close upper code-length limit for the AC codes, let alone to find a maximum AC code-length configuration.

In order to develop a strategy for deriving close upper limits of the AC code-lengths, we simplify the problem. We note that the AC coefficients cannot be disentangled from the DC coefficient in the Cube Condition (2) and the Integer Condition (3). In contrast, the AC Ball condition (1) makes no reference to the DC coefficient and is certainly much simpler than the system of 64 inequalities (2). For these reasons, we drop the Cube Condition (2) and the Integer Condition (3) and work with the AC Ball condition (1) alone. Thus instead of considering the configuration space $J$, we allow for configurations within the entire Ball $B$.

We only need to calculate an upper limit of the AC code-length for the entire Ball $B$, since such an upper limit is certainly also an upper limit of the code-length for any subset of $B$, in particular for the original configuration space $J$ and the sub-lattice defined by the Integer Condition. In fact, we only need to calculate an upper limit for configurations on the surface $S$ of the Ball $B$, since we have shown in the previous section that the code-length $len(F)$ takes its maximum value on the surface $S$. For this reason, we henceforth consider only configurations on the sphere $S$.

In what follows, we characterize a configuration $F$ on the sphere $S$ by its AC coefficients in the zigzag scan order, i.e. $F(k)$, $k=1,...,63$. In view of the AC Ball condition (1), the DC coefficient is fixed up to sign by the square root of $2^{20} - \sum_{(u,v) \neq (0,0)} (F_{uv})^2 > 0$.

Let $F$ be an arbitrary configuration on $S$. Let $Q(k)$, $k=1,...,63$, be the AC quantization values in the zigzag scan order. The code-length $len(F)$ of $F$ depends on each coefficient $F(k)$ only through the respective size $s_k$ of its quantized value $D(k) = \text{INT}(F(k)/Q(k))$, where $s_k$ is the smallest integer with the property $|D(k)| < 2^{s_k}$. Therefore, each coefficient $F(k)$ of $F$ can be replaced by the smallest possible non-negative value $Q(k) \times 2^{s_k - 1}$ when $s_k > 0$ or by zero when $s_k = 0$ without changing the code-length. Through these replacements the Huffman codes remain the same, and only the fixed-length codes for the amplitudes change. Each replaced coefficient is smaller than or equal to the original coefficient amplitude, so that the configuration $F$ still satisfies the AC Ball condition (1), which now reads as follows:

$$\sum_{k(s_k > 0)} Q(k)^2 \times 2^{2s_k - 2} < 2^{20}. \tag{6}$$

Note that the configuration $F$ still lies on the sphere $S$, although the DC coefficient may have changed. In order to calculate an upper code-length limit for configurations on $S$, we need to calculate an upper code-length limit only for such configurations $F$ having smallest non-negative coefficients. These configurations on $S$ shall be called *reduced configurations*. Unless specified otherwise, in the following all configurations are reduced configurations.

We now outline our strategy for calculating an upper code-length limit. We first choose a reference configuration $F_R$ on the sphere $S$. The reference $F_R$ cannot be just any configuration, but should have large coefficient sizes. We then compare $F_R$ with an arbitrary configuration $F$ on $S$ by introducing a set of operations which replace the coefficient sizes of $F_R$ by those of $F$. These operations generally correspond to the JPEG symbols $(r,s)$, so that a single operation may replace $r+1$ coefficients of $F_R$ to produce a run of $r$ zeros followed by a coefficient of (quantized) size $s$ in $F$. Rather than comparing $F_R$ and $F$ in terms of individual coefficients, we compare them in terms of symbols. This method enables a direct calculation of code-length gain or loss induced by each operation, since the Huffman codes are assigned to the symbols. Depending on $F$, some operations ("positive operations") enlarge certain coefficient sizes which generally lead to a longer code-length, whereas other operations ("negative operations") reduce some coefficient sizes or even introduce runs of zeros, which generally lessen the code-length. Since $F_R$ has already large coefficient sizes, there are only a limited number of positive operations possible at all. In general, not all operations can be positive, since otherwise the AC Ball condition (6) would be violated. Instead, positive operations must be accompanied by negative operations in order to fufill the AC Ball condition. We will derive an interdependence rule between the positive and negative operations from the AC Ball condition. By carefully choosing the reference $F_R$ (and introducing a few more conditions), this interdependence rule can be made very restrictive, such that only a few combinations of positive and negative operations are allowed. One of the key features of our method is the fact that for these limited number of combinations, an upper bound on the code-length gain can be calculated by a simple table look-up method. Finally, this upper bound is added to the code-length of the reference configuration $F_R$ to yield the upper limit on the AC code-length of configurations on the sphere $S$.

In order to specify the outlined strategy, it is helpful to straight away introduce an appropriate reference configuration $F_R$ and together with other necessary conditions which will lead to the restrictive interdependence rule. To this end, consider the AC Ball condition (6) and suppose for a moment that all quantization factors are powers of 2. Then, all AC coefficients of the configuration $F$ are either zero or powers of 2 since $F$ is a reduced configuration. We introduce the (unquantized) size $S(k)$ of each coefficient by defining $F(k) = 2^{S(k)-1}$ ($S(k) = 1,...,10$) if $F(k) > 0$ or by setting $S(k) = 0$ if $F(k) = 0$. For indices $i = 1,2,...,a$, let $p_i$ be positions in the zigzag order with the property $F(p_i) = 2^8$, i.e., $S(p_i) = 9$. Likewise, for $j = 1,2,...,b$, let $q_j$ be positions such that $F(q_j) = 2^9$ ($S(q_j) = 10$). The AC Ball condition (6) can now be converted into the following inequality:

$$\sum_{k(S(k)>0)} 2^{2S(k)-2} < 2^{20} \qquad \Leftrightarrow$$

$$\sum_{i=1}^{a} 2^{16} + \sum_{j=1}^{b} 2^{18} + \sum_{k \neq p_i, k \neq q_j, S(k)>0} 2^{2S(k)-2} < 2^{20} \qquad \Leftrightarrow$$

$$\sum_{k \neq p_i, k \neq q_j, S(k)>0} 2^{2S(k)-2} < (64 - 4a - 16b) \cdot 2^{14}. \qquad (7)$$

This inequality implies the constraint $a + 4b < 16$, which restricts the numbers of size 9 and size 10 indices to just 40 combinations of the pair $(a,b)$. In particular, $0 \leq a \leq 15$ and $0 \leq b \leq 3$. The right hand side is a sum of $64 - 4a - 16b$ copies of $2^{14}$, whereas the left hand side sums over at most $63 - a - b$ indices $k$. Hence we conclude that at least $3a + 15b$ terms on the left hand side must be smaller than $2^{14}$. These terms have respective sizes $S(k)$ less than 8. The maximum value for the expression $3a + 15b$ is 54.

These observations suggest a natural choice for the reference configuration $F_R$, namely to set each of its coefficients to the same value $F_R(k) = 2^7$ with size 8. If $a$ ($b$) coefficients of $F_R$ are replaced by size 9 (10) coefficients ("positive operations"), at least $3a$ ($15b$) coefficients of $F_R$ must be replaced by smaller sized coefficients ("negative operations"). Altogether, no more than 15 coefficients can be replaced by size 9 coefficients (with amplitude $2^8$) and no more than 3 coefficients can be replaced by size 10 coefficients, since otherwise the AC Ball condition is violated no matter how small the other coefficients might be.

The necessary condition for this restrictive interdependence rule is that the quantization values be powers of 2. In order to derive the same rule for an arbitrary quantization table $Q(k)$, we replace it by the "power-of-2 quantization table" defined as follows:

$$Q_2(k) := 2^{C(k)}, \text{ where } C(k) = \text{INT}(\log_2 Q(k)) \quad (\text{"Power-of-2 Q-Table"}). \tag{8}$$

We are allowed to make this replacement, since we only need to find an upper limit of AC code-length for this power-of-2 quantization table $Q_2(k)$ rather than for the original quantization table $Q(k)$. This is true in view of the following proposition:

<u>Proposition 3.</u> *On the sphere $S$, the maximum code-length for $Q_2(k)$ is larger than or equal to the maximum code-length for the original quantization table $Q(k)$.*

*Proof.* To prove this assertion, let $F$ be a maximum code-length configuration on $S$ (based on the quantization table $Q(k)$) satisfying the AC Ball condition (6). Define a new configuration $F'$ on $S$ by $F'(k) := Q_2(k)/Q(k) \times F(k)$. Since $Q_2(k) \leq Q(k)$, the new configuration $F'$ still fulfills the AC Ball condition (6), wherein each $Q(k)$ is replaced by $Q_2(k)$. By construction, quantizing $F(k)$ with $Q(k)$ yields the same result as when quantizing $F'(k)$ with $Q_2(k)$, so that the code-length of $F$ is the same as the code-length of $F'$ based on $Q_2(k)$. By the very definition of the maximum code-length for $Q_2(k)$, it is greater than or equal to this code-length of $F'$, which proves the assertion.

This assertion is also true for the entire Ball $B$, but might not be true for the rotated cube $J$. The difficulties involved in the above reasoning can be visualized without going into the details of the inequalities (2). Let $F$ be a configuration at or near a vertex of the cube. If the cube was not rotated, reducing coefficient amplitudes of $F$, e.g. in order to define $F'$, will move $F$ entirely within the cube. However, for a rotated cube like $J$, reductions of some of the amplitudes might move $F$ along a coordinate direction outside the cube, so that the new configuration $F'$ may not be well-defined. Although this does not disprove the assertion, we see that it is not straightforward to introduce the concept of the power-of-2 quantization table (8) without enlarging the configuration space $J$ to the Ball $B$.

In summary, the arguments set forth in this section and Proposition 3 simplifies the problem of finding an upper AC code-length limit in the following way:

Theorem 1. *For a given quantization table $Q(k)$, the AC code-length of any DCT configuration in the configuration space $J$ is less than or equal to the maximum code-length of reduced configurations on the sphere $S$ using the power-of-2 quantization table $Q_2(k)$ based on $Q(k)$. Therefore, an upper code-length limit for such reduced configurations is also an upper code-length limit of all DCT configurations.*

The introduction of a power-of-2 quantization table via (8) is not mandatory in deriving a restrictive interdependence rule. However, for general quantization factors $Q(k)$, it seems that much more work is needed to derive from the AC Ball condition (6) a convenient interdependence rule. For example, the quantized coefficient sizes $s_k$ of the reference may be defined as $s_k = \text{INT}(8 - \log_2 Q(k))$, so that each square term in (6) comes close to $2^{14}$. However, since these reference values may not be constant over the positions $k$, it is very difficult to derive a simple interdependence rule in the manner shown above. In the rest of this work we restrict ourselves to the power-of-2 quantization table restriction (8). We leave the study of the general case to future work.

The unquantized sizes $S(k)$ determine the coefficients $F(k)$ of a configuration $F$. The quantization with a power-of-2 quantization table amounts to reducing these sizes by the exponents $C(k)$ of the quantization factors. The quantized sizes $s_k$ are given by $s_k = \max(S(k) - C(k), 0)$. Since the coefficients $F(k)$ are chosen to be minimal, $S(k)$ is zero if $s_k$ is zero.

## IV. DETAILED STRATEGY FOR CALCULATING THE UPPER LIMIT

Following the outline of the strategy, let $F_R$ be the reference configuration on $S$ having coefficients with the constant value $F_R(k) = 2^7$ and size $S_R(k) = 8$. The quantized sizes are given by $\bar{s}_k = \max(S_R(k) - C(k), 0)$. Since the largest quantization factor in this work is $121$, the largest power-of-2 quantization factor is $64 = 2^6 < 121$. Thus the quantized sizes of $F_R$ are always larger than or equal to $2$, $\bar{s}_k \geq 2$. In particular, all coefficients of $F_R$ are non-zero after quantization.

Let $F$ be an arbitrary configuration on $S$ having unquantized sizes $S(k)$ and quantized sizes $s_k = \max(S(k) - C(k), 0)$. We now define the following operations OP1 to OP6, which generally correspond to symbols $(r, s)$. Depending on the size $S(p)$ of the target configuration $F$, a different operation replaces the size $S_R(p)$ of the reference configuration $F_R$ by the size $S(p)$ of $F$:

OP1:　If $0 < S(p) < S_R(p)$ and $S(p-1) > 0$ or $p = 1$, replace $S_R(p) \mapsto S(p)$.

OP2:　If $0 < S(p) < S_R(p)$, $S(p-1) = S(p-2) = ... = S(p-r) = 0$ for $r > 0$, and $S(p-r-1) > 0$ or $p - r = 1$, replace simultaneously $r+1$ sizes $S_R(p-r), ..., S_R(p) \mapsto S(p-r), ..., S(p-1), S(p)$.

OP3:　If $S(p) \geq S_R(p)$, $S(p-1) = S(p-2) = ... = S(p-r) = 0$ for $r > 0$, and $S(p-r-1) > 0$ or $p - r = 1$, replace simultaneously $r$ sizes $S_R(p-r), ..., S_R(p-1) \mapsto S(p-r), ..., S(p-1)$.

OP4:　If $S(p) > 0$ and $S(q) = 0$ for all $q > p$, replace $S_R(q) \mapsto S(q)$ for all $q > p$ simultaneously.

OP5:　If $S(p) > S_R(p)$ and $S(p-1) > 0$ or $p = 1$, replace $S_R(p) \mapsto S(p)$.

OP6:　If $S(p) > S_R(p)$, $S(p-1) = S(p-2) = ... = S(p-r) = 0$ for $r > 0$, and $S(p-r-1) > 0$ or $p - r = 1$, then, after performing OP3, replace $S_R(p) \mapsto S(p)$.

Proposition 4. *Given a reduced configuration $F$ for a power-of-2 quantization table, the set of operations OP1 to PO6 is uniquely defined.*

*Proof.* It suffices to show that each coefficient $F_R(p)$ can be replaced by the corresponding coefficient $F(p)$ through exactly one of these operations if $F_R(p) \neq F(p)$. We consider several cases of $F(p)$. The coefficient size $S(p)$ can be either zero or non-zero. If $S(p)$ is non-zero and not preceded by zeros, three cases may arise: if $S(p) = S_R(p)$, we see that none of the operations OP1 to OP6 can be applied at all; if $S(p) < S_R(p)$, then $S_R(p)$ is replaced by $S(p)$ only through OP1; if $S(p) > S_R(p)$, then $S_R(p)$ is replaced by $S(p)$ only through OP5. On the other hand, if the non-zero $S(p)$ is preceded by a run of $r$ zeros, again three cases may arise: if $S(p) = S_R(p)$, then $r$ preceeding coefficient sizes $S_R(p-r),...,S_R(p-1)$ are replaced by zeros through OP3, but OP3 does not change $S_R(p)$ itself; if $S(p) < S_R(p)$, then $r+1$ coefficient sizes $S_R(p-r),...,S_R(p)$ are replaced by the respective sizes of $F$ through OP2; if $S(p) > S_R(p)$, then $r$ coefficient sizes $S_R(p-r),...,S_R(p-1)$ are replaced by zeros through OP3, whereupon the coefficient $S_R(p)$ is replaced by $S(p)$ through OP6. Now assume that $S(p)$ is zero. Then $S(p)$ is either a member of a zero run followed by a non-zero coefficient or it is a member of a zero run including the last zigzag coefficient. In the first case, $S_R(p)$ will be replaced by zero through either OP2 or OP3, whereas in the latter case, $S_R(p)$ will be replaced by zero through OP4. In summary, each coefficient $F_R(p)$ of $F_R$ is correctly replaced by the respective coefficient $F(p)$ of $F$ by one and only one operation from OP1 to OP6.

Following the outline of our strategy, we examine the changes in the AC code-length induced by each of the operations. To this end, we introduce the notion of *local (or individual) code-length difference* $\delta$, which is defined to be the change in code-length per affected position: if an operation replaces sizes at $m$ indices (or positions) $p_i$ and leaves sizes at $n$ other indices unaffected, then the local code-length change is defined to be $\delta = \delta(t_i, u_j) = (len(\bar{t}_i, u_j) - len(t_i, u_j))/m$, where $t_i$ are the quantized sizes at positions $p_i$, $\bar{t}_i$ are the corresponding replaced quantized sizes, and $u_j$ are the unaffected quantized sizes at $n$ positions. We assign to each of the $m$ affected positions $p_i$ the code-length difference $\delta$. The differences at these $m$ positions together yield the total change in code-length through the operation. For example, OP2 replaces $r+1$ coefficients simultaneously, so that $m = r+1$ and $n = 0$. As for OP3, it replaces $r$ coefficients, so that $m = r$. Since it leave the size at $p$ unaffected, we have $n = 1$. For OP6, we have $m = 1$ and $n = r$.

Proposition 5. *The operations OP1 to OP4 reduce the code-length ("negative operations"), i.e., the respective local code-length differences are negative or zero. The operations OP5 and OP6 increase the code-length ("positive operations"), whereby the local code-length differences are strictly positive.*

*Proof.* We consider each operation in succession.

OP1 reduces the unquantized size $S_R(p) = 8$ to a lower size $S(p)$ and so induces a change of the quantized size $\bar{s}_p$ to a smaller size $s_p$. The size $s_p$ is non-zero, since otherwise $S(p)$ must be zero, so that OP1 would not have been applied. As noted earlier, for any runlength $r \geq 0$ and for any sizes $s > s' > 0$, the relation $len(r, s) > len(r, s')$ holds. The local code-length difference is given by $-\delta_{p,s}^1 = len(0, s) - len(0, \bar{s}_p) < 0$, where we have set $s = s_p$ to simplify the notation. Thus OP1 is a negative operation, which reduces the AC code-length.

As for OP2, we first note that for the example quantization tables K.1 and K.2 of the JPEG specification, preceding quantization factors in the zigzag scan order are always smaller than twice the current quantization factor, i.e., $Q_0(l) < 2Q_0(k)$ if $l < k$. After scaling the quantization factors according to (4), we still have a weaker relation $Q(l) \leq 2Q(k) + 1$. For the exponents of the power-of-2 quantization factors defined according to (8), we can deduce the relation $C(l) \leq C(k) + 1$ by using the fact that the quantization factors are integers. For the quantized sizes in the reference configuration $F_R$ this relation implies $\bar{s}_{p-r}, ..., \bar{s}_{p-1} \geq \bar{s}_p - 1 > 0$. It follows that

$$len(\bar{s}_{p-r}, \cdots \bar{s}_{p-1}, \bar{s}_p) = len(0, \bar{s}_{p-r}) + \cdots + len(0, \bar{s}_p) \geq r \cdot len(0, \bar{s}_p - 1) + len(0, \bar{s}_p).$$

Tables I and II below reveal that the right hand side expression is always greater than or equal to $len(r, \bar{s}_p)$, which is greater than $len(r, s_p)$ since $\bar{s}_p > s_p$. In summary, $len(\bar{s}_{p-r}, \cdots \bar{s}_{p-1}, \bar{s}_p) > len(r, s_p)$, which was to be shown. Let $-\delta_{p,r,s}^2 = (len(r, s) - len(\bar{s}_{p-r}, \cdots \bar{s}_{p-1}, \bar{s}_p))/(r+1) < 0$ be the local code-length difference assigned to each of the $r+1$ positions $p-r$ to $p$, where we have set $s = s_p$.

OP3 differs from OP2 in that the size at position $p$ is not changed. Repeating the arguments set forth for OP2, we see that the code-length is reduced or remains the same. Let $-\delta_{p,r}^3 = (len(r, \bar{s}_p) - len(\bar{s}_{p-r}, \cdots \bar{s}_{p-1}, \bar{s}_p))/r \leq 0$ denote the local code-length difference assigned to each of the $r$ positions $p-r$ to $p-1$.

As for OP4, we consider the inequality $len(\bar{s}_{p+1}, ..., \bar{s}_{63}) \geq (63 - p) \cdot len(0, 2)$, which is true since quantized sizes of the reference configuration are greater than or equal to $2$. For luminance AC coeffi-cients, we have the equality $len(0,2) = 4 = len(EOB)$, whereas $len(0,2) = 5 > 2 = len(EOB)$ for chrominance AC coefficients. Thus OP3 always reduces the code-length except when $p = 62$ and $\bar{s}_{63} = 2$ for luminance AC coefficients, in which case there is no change in the code-length. Let $-\delta_p^4 = (len(EOB) - len(\bar{s}_{p+1}, \cdots, \bar{s}_{63}))/(63-p) \leq 0$ be the local code-length difference assigned to each of the affected $63 - p$ positions.

OP5 enlarges the size $\bar{s}_p$ to either $s_p = \bar{s}_p + 1$ or to $s_p = \bar{s}_p + 2$. These are the only possible cases, since for the unquantized size $S(p)$ of the target configuration $F$ only two sizes 9 and 10 are available above the unquantized size $S_R(p) = 8$. Let $\alpha_p^1 = len(0, \bar{s}_p + 1) - len(0, \bar{s}_p) > 0$ be the local code-length gain for the case $s_p = \bar{s}_p + 1$, and similarly let $\beta_p^1 = len(0, \bar{s}_p + 2) - len(0, \bar{s}_p) > 0$ be the local code-length gain for the case $s_p = \bar{s}_p + 2$.

OP6 induces a code-length gain of $len(r, s_p) - len(r, \bar{s}_p) > 0$. Similar to the case of OP5, only the two cases $s_p = \bar{s}_p + 1$ or $s_p = \bar{s}_p + 2$ may occur. Let $\alpha_{p,r}^2 = len(r, \bar{s}_p + 1) - len(r, \bar{s}_p) > 0$ be the local code-length gain for $s_p = \bar{s}_p + 1$, and similarly let $\beta_{p,r}^2 = len(r, \bar{s}_p + 2) - len(r, \bar{s}_p) > 0$ be the local code-length gain for $s_p = \bar{s}_p + 2$.

In the above proof, we have introduced notations for the local code-length losses and gains: $-\delta_{p,s}^1$, $-\delta_{p,r,s}^2$, $-\delta_{p,r}^3$, $-\delta_p^4$, $\alpha_p^1$, $\beta_p^1$, $\alpha_{p,r}^2$, and $\beta_{p,r}^2$. Note that the indices assigned for each code-length change uniquely determine the respective operation. For example, $-\delta_{p,s}^1$ implies that OP1 is applied to a coefficient at zigzag scan position $p$ where the quantized size of the configuration $F$ is given by $s$. OP3 and OP4 do not depend on the quantized size, so that their code-length losses are indexed without the size.

The code-length of $F$ can be represented as a sum of the code-length of $F_R$ and the local code-length differences for all positions affected by the operations OP1 to OP6. With the help of the indices for the local code-length differences, we can express the code-length of $F$ as follows:

$$len(F) = len(F_R) + A + B - \Delta , \text{ wherein} \tag{9}$$

$$A = \alpha_{q_1}^1 + ... + \alpha_{q_{a'}}^1 + \alpha_{q_{a'+1}, r_{a'+1}}^2 + ... + \alpha_{q_a, r_a}^2 , \tag{10}$$

$$B = \beta^1_{\bar{q}_1} + \ldots + \beta^1_{\bar{q}_{b'}} + \beta^2_{\bar{q}_{b'+1}, \bar{r}_{b'+1}} + \ldots + \beta^2_{\bar{q}_b, \bar{r}_b}, \text{ and} \tag{11}$$

$$\begin{aligned}
\Delta = & \; 1 \cdot \delta^1_{p_1, s_1} + 1 \cdot \delta^1_{p_2, s_2} + \ldots \\
& + (r'_1 + 1) \cdot \delta^2_{p'_1, r'_1, s'_1} + (r'_2 + 1) \cdot \delta^2_{p'_2, r'_2, s'_2} + \ldots \\
& + (r''_1) \cdot \delta^3_{p''_1, r''_1, s''_1} + (r''_2) \cdot \delta^3_{p''_2, r''_2, s''_2} + \ldots \\
& + (63 - p) \cdot \delta^4_p.
\end{aligned} \tag{12}$$

The sum $A$ counts the local code-length gains induced by the positive operations OP5 and OP6 for the case of unquantized size 9, whereas the sum $B$ counts the local code-length gains for the case of unquantized size 10. As indicated by the indices, we assume that there are $a$ terms in $A$ and $b$ terms in $B$. Thus the configuration $F$ is assumed to posses $a$ positions having unquantized size 9 and $b$ positions with unquantized size 10.

Due to the AC Ball condition (7), there must exist at least $3a + 15b$ positions in $F$ with unquantized size less than 8. Each such position must have been created by one of the negative operations OP1 to OP4, i.e. it is either a position within a zero run or a position having non-zero size smaller than 8. The local code-length losses of these $3a + 15b$ or more positions are summarized in the sum $\Delta$. It has contributions from four lines on the right-hand side of equation (12), which are the sums of code-length losses generated by the negative operations OP1, OP2, OP3, and OP4, respectively.

We now estimate an upper code-length limit for the expression (9). To this end, we observe that the sum $\Delta$ of local code-length losses in (12) has been generated by operations OP1 to OP4 uniquely defined for the particular configuration $F$. Therefore, the loss $\Delta$ is certainly greater than or equal to the sum $\Delta(3a + 15b)$, which is defined to be the sum of only the $3a + 15b$ smallest local code-length losses among any and all possible local code-length losses created by any and all possible operations OP1 to OP4 for any and all target configurations. More precisely, let $\Delta\{\}$ be the set of all local code-length losses $\delta^1_{p,s}$ for all possible pairs $(p, s)$, $r + 1$ copies of $\delta^2_{p,r,s}$ for all possible triplets $(p, r, s)$, $r$ copies of $\delta^3_{p,r}$ for all possible pairs $(p, r)$, and $(63 - p)$ copies of $\delta^4_p$ for all positions $p$. Define the *loss function* $\Delta(n)$ to be the sum of the $n$ smallest values in $\Delta\{\}$. Then the relation $\Delta \geq \Delta(3a + 15b)$ holds.

In a similar fashion, let $A\{\}$ be the set of all local code-length gains $\alpha_p^1$ for all positions $p$ and all $\alpha_{p,r}^2$ for all possible pairs $(p,r)$ for any and all target configurations. Likewise, let $B\{\}$ be the set of all local code-length gains $\beta_p^1$ for all positions $p$ and all $\beta_{p,r}^2$ for all possible pairs $(p,r)$ for any and all target configurations. Define the *gain function* $A(n)$ to be the sum of the $n$ largest values in the set $A\{\}$, and the *gain function* $B(n)$ to be the sum of the $n$ largest values in $B\{\}$. Clearly then, the relations $A \leq A(a)$ and $B \leq B(b)$ hold. In summary we obtain the inequality

$$len(F) \leq len(F_R) + \big(A(a) + B(b) - \Delta(3a + 15b)\big). \tag{13}$$

The second term on the right-hand side is certainly not greater than the maximum of the same expression $A(a) + B(b) - \Delta(3a + 15b)$ among all $40$ possible combinations of the pair $(a,b)$. Thus, we have just proved the following:

Theorem 2. *Given an arbitrary reduced configuration $F$, its code-length satisfies the inequality*

$$len(F) \leq \Lambda, \text{ where} \tag{14}$$

$$\Lambda = len(F_R) + \max_{(a,b)}\{A(a) + B(b) - \Delta(3a + 15b)\}. \tag{15}$$

*The formula (15) represents an upper limit of the AC code-length for reduced configurations on the sphere $S$ when using a power-of-2 quantization table. By Theorem 1, $\Lambda$ is an upper code-length limit for all DCT configurations.*

The code-length $len(F_R)$ can be readily calculated by using the power-of-2 quantization values $Q_2(k)$ in (8). In order to calculate the maximum term in (15), we have to determine the $15$ largest local code-length gains in the set $A\{\}$, the $3$ largest local code-length gains in the set $B\{\}$, and the $54$ smallest local code-length losses in the set $\Delta\{\}$. In principle, these local code-length changes can be determined by applying each of the operations OP1 to OP6 for all possible parameter values of $r$ and $s$ at all possible positions $p$ in $F_R$ and calculating the induced code-length changes: for the code-length loss $-\delta_{p,s}^1$, there are 63 possible positions $p$ and 7 possible non-zero sizes $s < 8$, so that we may have to calculate the loss values for up to $63 \times 7$ cases; for $-\delta_{p,r,s}^2$, there are at most $(63 \times 62)/2 \times 7$ cases to calculate, where we accounted for the fact that $r < p$; for $-\delta_{p,r}^3$, there are at most $(63 \times 62)/2$ cases; for $-\delta_p^4$, there are at most 62 cases; for $\alpha_p^1$ and $\beta_p^1$, there are at most 63 cases; for $\alpha_{p,r}^2$ and $\beta_{p,r}^2$, there are at most $(63 \times 62)/2$ cases. By adding all these numbers, we obtain 20159 calculations, which can be easily done by a computer. This relatively small number of calculations is sufficient to completely determine the loss and gain functions. On the other hand, if we had to consider the combinations of the operations for all possible configurations $F$, the calculations of the code-length changes would have been impossible.

It turns out that only a few combinations of the parameter values need to be considered to determine the loss and gain functions in the required ranges. Tables I and II reveal that $len(r,s) > len(r,s')$ for $r \geq 0$ and $s > s' > 0$ and $len(r,s) = len(r',s)$ for large values of $r$ and $r'$. Using these relations, calculations for large runlength values and large/small sizes may be skipped. In this way it is possible to calculate the upper limit value by hand, see the calculation example in the next section.

We can easily improve the upper limit $\Lambda$. To see how, let us assume that the quantized coefficient sizes $\bar{s}_p$ of $F_R$ have the same value when $p = 60, 61, 62, 63$. Since the sizes are the same, the local code-length losses $\delta_{61,1,1}^2$, $\delta_{62,1,1}^2$, and $\delta_{63,1,1}^2$ in the set $\Delta\{\}$ have the same value, say $\delta$. There are altogether 6 copies of $\delta$ in $\Delta\{\}$, since each of the underlying operations OP2 affects two positions. If $\delta$ was among the smallest values in $\Delta\{\}$, the loss function $\Delta(n)$ would add $\delta$ up to 6 times. On the other hand, there are only 4 positions $p = 60, 61, 62, 63$ available for assigning $\delta$, so that 2 copies of $\delta$ are superfluous in $\Delta\{\}$. In other words, no configuration can have more than 4 times the local loss $\delta$. By eliminating such superfluous copies in the set $\Delta\{\}$, we obtain a subset $\widetilde{\Delta}\{\}$. In practical application, it is difficult to find and eliminate all superfluous copies, so that we may delete some of such copies to obtain the subset $\widetilde{\Delta}\{\}$. The term $\Delta$ summarizes the local code-length losses for a real configuration $F$, so that it must be still greater than or equal to the sum of the $3a + 15b$ smallest local code-length losses within this subset $\widetilde{\Delta}\{\}$, wherein superfluous copies have been eliminated. Since $\tilde{\Delta}\{\}$ is a subset of $\Delta\{\}$, the sum $\tilde{\Delta}(n)$ of the $n$ smallest values in $\tilde{\Delta}\{\}$ is greater than or equal to $\Delta(n)$, so that we obtain the relation $\Delta \geq \tilde{\Delta}(3a+15b) \geq \Delta(3a+15b)$. Arguing in a similar fashion, we can eliminate (at least some) superfluous counts of local code-length gains in the sets $A\{\}$ and $B\{\}$ to obtain respective subsets $\tilde{A}\{\}$ and $\tilde{B}\{\}$. Let $\tilde{A}(n)$ and $\tilde{B}(n)$ be the sums of the $n$ smallest local code-length gains in $\tilde{A}\{\}$ and $\tilde{B}\{\}$, respectively, for which the relations $A \leq \tilde{A}(a) \leq A(a)$ and $B \leq \tilde{B}(b) \leq B(b)$ are valid. In summary, we obtain a new, possibly lower upper limit

$$\tilde{\Lambda} = len(F_R) + \max_{(a,b)} \left\{ \tilde{A}(a) + \tilde{B}(b) - \tilde{\Delta}(3a+15b) \right\} \leq \Lambda . \tag{16}$$

To further improve this upper limit, we note that the upper limit needs to be calculated for maximum code-length configurations only. Obviously, the sum $\Delta$ of a maximum code-length configuration must be greater than or equal to the sum of $3a+15b$ smallest local code-length losses for all possible maximum code-length configurations. Therefore we can eliminate from the subset $\tilde{\Delta}\{\}$ those losses which do not arise for a maximum code-length configuration, and obtain an improved loss function $\tilde{\tilde{\Delta}}(n) \geq \tilde{\Delta}(n)$ with the property $\Delta \geq \tilde{\tilde{\Delta}}(3a+15b) \geq \tilde{\Delta}(3a+15b)$. Similarly, by eliminating (at least some) gains which can not occur for a maximum code-length configuration from the sets $\tilde{A}\{\}$ and $\tilde{B}\{\}$, we obtain smaller gain functions $\tilde{\tilde{A}}(n)$ and $\tilde{\tilde{B}}(n)$ with the properties $A \leq \tilde{\tilde{A}}(a) \leq \tilde{A}(a)$ and $B \leq \tilde{\tilde{B}}(b) \leq \tilde{B}(b)$. In summary, we obtain a new, possibly lower upper limit $\tilde{\tilde{\Lambda}} \leq \tilde{\Lambda}$.

To find those local code-length changes which do not occur in a maximum code-length configuration, consider the operations OP2, OP3, and OP6. These operations generate runs of zeros in $F$, say $0,...,0,S(p)$. Let $(r, s_p)$ be the symbol after quantization, $len(r, s_p)$ its code-length, and assume $s_p > 2$. We compare the sub-sequence $0,...,0,S(p)$ with another possible sub-sequence of $r+1$ coefficients at the same positions, wherein $S(p)$ is replaced by the value $S(p)-1$ and at most 3 preceding zeros are replaced by the same non-zero value $S(p)-1$. Such replacements within the configuration $F$ are allowed, since the replaced configuration still satisfies the AC Ball condition (5) due to the identity $2^{2S(p)-2} = 4 \cdot 2^{2(S(p)-1)-2}$. In view of the relation $C(l) \leq C(k)+1$ ($l < k$) for the exponents of the power-of-2 quantization factors, the new sub-sequence has up to 3 non-zero quantized sizes greater than or equal to $s_p - 2$, followed by $s_p - 1$. If the code-length $len(r, s_p)$ is smaller than the code-length of this replaced sub-sequence, the original sequence $0,...,0,S(p)$ cannot occur in a maximum code-length configuration, since it is replaceable with a sub-sequence with a longer code-length. Tables I and II summarize the code-lengths $len(r,s)$ of zero run sub-sequences. Runlength/size combinations of shaded cells in Tables I and II cannot appear in a maximum code-length configuration.

Table I: Code-lengths (Huffman + size) for chrominance AC coefficients

| Size | Runlength | | | | | | | | | | | | | | | |
|---|---|---|---|---|---|---|---|---|---|---|---|---|---|---|---|---|
| | 0 | 1 | 2 | 3 | 4 | 5 | 6 | 7 | 8 | 9 | 10 | 11 | 12 | 13 | 14 | 15 |
| 0 | 2 | | | | | | | | | | | | | | | 10 |
| 1 | 3 | 5 | 6 | 6 | 7 | 7 | 8 | 8 | 9 | 10 | 10 | 10 | 10 | 12 | 15 | 16 |
| 2 | 5 | 8 | 10 | 10 | 11 | 12 | 13 | 13 | 18 | 18 | 18 | 18 | 18 | 18 | 18 | 18 |
| 3 | 7 | 11 | 13 | 13 | 19 | 19 | 19 | 19 | 19 | 19 | 19 | 19 | 19 | 19 | 19 | 19 |
| 4 | 9 | 13 | 16 | 16 | 20 | 20 | 20 | 20 | 20 | 20 | 20 | 20 | 20 | 20 | 20 | 20 |
| 5 | 10 | 16 | 20 | 21 | 21 | 21 | 21 | 21 | 21 | 21 | 21 | 21 | 21 | 21 | 21 | 21 |
| 6 | 12 | 18 | 22 | 22 | 22 | 22 | 22 | 22 | 22 | 22 | 22 | 22 | 22 | 22 | 22 | 22 |
| 7 | 14 | 23 | 23 | 23 | 23 | 23 | 23 | 23 | 23 | 23 | 23 | 23 | 23 | 23 | 23 | 23 |
| 8 | 17 | 24 | 24 | 24 | 24 | 24 | 24 | 24 | 24 | 24 | 24 | 24 | 24 | 24 | 24 | 24 |
| 9 | 19 | 25 | 25 | 25 | 25 | 25 | 25 | 25 | 25 | 25 | 25 | 25 | 25 | 25 | 25 | 25 |
| 10 | 22 | 26 | 26 | 26 | 26 | 26 | 26 | 26 | 26 | 26 | 26 | 26 | 26 | 26 | 26 | 26 |

Table II: Code-lengths (Huffman + size) for luminance AC coefficients

| Size | Runlength | | | | | | | | | | | | | | | |
|---|---|---|---|---|---|---|---|---|---|---|---|---|---|---|---|---|
| | 0 | 1 | 2 | 3 | 4 | 5 | 6 | 7 | 8 | 9 | 10 | 11 | 12 | 13 | 14 | 15 |
| 0 | 4 | | | | | | | | | | | | | | | 11 |
| 1 | 3 | 5 | 6 | 7 | 7 | 8 | 8 | 9 | 10 | 10 | 10 | 11 | 11 | 12 | 17 | 17 |
| 2 | 4 | 7 | 10 | 11 | 12 | 13 | 14 | 14 | 17 | 18 | 18 | 18 | 18 | 18 | 18 | 18 |
| 3 | 6 | 10 | 13 | 15 | 19 | 19 | 19 | 19 | 19 | 19 | 19 | 19 | 19 | 19 | 19 | 19 |
| 4 | 8 | 13 | 16 | 20 | 20 | 20 | 20 | 20 | 20 | 20 | 20 | 20 | 20 | 20 | 20 | 20 |
| 5 | 10 | 16 | 21 | 21 | 21 | 21 | 21 | 21 | 21 | 21 | 21 | 21 | 21 | 21 | 21 | 21 |
| 6 | 13 | 22 | 22 | 22 | 22 | 22 | 22 | 22 | 22 | 22 | 22 | 22 | 22 | 22 | 22 | 22 |
| 7 | 15 | 23 | 23 | 23 | 23 | 23 | 23 | 23 | 23 | 23 | 23 | 23 | 23 | 23 | 23 | 23 |
| 8 | 18 | 24 | 24 | 24 | 24 | 24 | 24 | 24 | 24 | 24 | 24 | 24 | 24 | 24 | 24 | 24 |
| 9 | 25 | 25 | 25 | 25 | 25 | 25 | 25 | 25 | 25 | 25 | 25 | 25 | 25 | 25 | 25 | 25 |
| 10 | 26 | 26 | 26 | 26 | 26 | 26 | 26 | 26 | 26 | 26 | 26 | 26 | 26 | 26 | 26 | 26 |

## V. CALCULATION AND RESULTS

We demonstrate the calculation of the upper limit $\Lambda$ (15) for chrominance AC coefficients when using the unscaled ($SF = 1$) quantization Table K.1. Its values are given in zigzag scan order: $Q(k) = 18$, 18, 24, 21, 24, 47, 26, 26, 47, 99, 66, 56, 66, 99,..., 99 for $k = 1,...,63$.

1. The power-of-2 quantization table is calculated according to (8). Its exponents are given by:

$C(k) = 4$ for $k = 1,2,3,4,5,7,8$, $C(k) = 5$ for $k = 6,9,12$, and $C(k) = 6$ for all other positions.

2. The quantized sizes of the reference $F_R$ are given by $\bar{s}_k = 4$ for $k = 1,2,3,4,5,7,8$, $\bar{s}_k = 3$ for $k = 6,9,12$, and $\bar{s}_k = 2$ for the remaining 53 positions. With the help of Table I, we obtain $len(F_R) = 7 \times 9 + 3 \times 7 + 53 \times 5 = 349$.

3. OP1: if we set $s := \bar{s}_p - 1$, then the local code-length losses at any of the 63 positions $p$ have the same value $\delta^1_{p,s} = len(0,\bar{s}_p) - len(0,\bar{s}_p - 1) = 2$. Any smaller choice of $s$ leads to a greater loss $\delta^1_{p,s} > 2$. Since we know already that the set $\Delta\{\}$ contains more than 54 copies of 2, losses greater than or equal to 2 need not be considered anymore.

4. OP2: it is always $\delta^2_{p,r,s} \geq 2$. For example, the replacement of the quantized sizes $2,2 \to 0,1$ yields a local loss of $\delta^2_{p,1,1} = (len(0,2) + len(0,2) - len(1,1))/2 = ((5+5) - 5)/2 = 2.5$. Another example is $\delta^2_{7,1,3} = (len(0,3) + len(0,4) - len(1,3))/2 = ((7+9) - 11)/2 = 2.5$ at position $p = 7$.

5. OP3: at position $p = 12$, we have $\delta^3_{12,1} = (len(0,2) + len(0,3) - len(1,3))/1 = 1$. For any other choices of $p$ and $r$, a simple lookup of Table I yields the result $\delta^3_{p,r} > 2$.

6. OP4: if $\bar{s}_{63} = 2$ is replaced by zero, we obtain $\delta^4_{62} = len(0,2) - len(EOB) = 3 > 2$. The more non-zero coefficients we replace by zeros, the larger the local code-length losses.

7. OP5: for all positions $p$, we have $\alpha^1_p \leq 2$. For example, $\alpha^1_{45} = len(0,3) - len(0,2) = 2$. Similarly, we obtain the result $\beta^1_p \leq 4$ for all $p$.

8. OP6: for positions $p = 10,11,13,14,...,63$ and runlength $r = 1,2$, we obtain $\alpha^2_{p,r} = 3$. For example, $\alpha^2_{16,2} = len(2,3) - len(2,2) = 13 - 10 = 3$. For all other combinations of $p$ and $r$, $\alpha^2_{p,r} \leq 3$. Likewise, $\beta^2_{p,2} = 6$ for $p = 10,11,13,14,...,63$, and $\beta^2_{p,r} < 6$ for all other $p$ and $r$.

9. Summary: the 54 smallest local code-length losses are one copy of 1 and 53 copies of 2, so that the loss function is given by $\Delta(n) = 2n - 1$ when $1 \leq n \leq 54$ and $\Delta(0) = 0$. If we consider (unquantized) size 9 coefficients, the 15 largest local code-length gains are 15 times the same value 3, so that the gain function is given by $A(a) = 3a$ ($a \leq 15$). Similarly, for unquantized size 10 we obtain the gain function $B(b) = 6b$ ($b \leq 3$). The total function $A(a) + B(b) - \Delta(3a + 15b)$ is zero when $a = b = 0$ and equals $-3a - 24b + 1$ when $a > 0$ or $b > 0$. Its maximum lies at $a = b = 0$, so that the upper code-length limit for chrominance AC coefficients is just given by the length of the reference configuration, $\Lambda = len(F_R) = 349$. This result means that the reference $F_R$ is in fact a maximum code-length configuration.

Upper limits for scale factors other than $SF = 1$ and for luminance AC coefficients can be calculated in a similar fashion. Table III shows calculation results for a variety of scale factors.

Table III: Upper limits of AC code-lengths in bits for different scale factors

| Scale factor | 1/64 | 1/16 | 1/8 | 1/6 | 1/4 | 1/2 | 1/1 |
|---|---|---|---|---|---|---|---|
| Luminance AC | 1134 | 956 | 812 | 715 | 654 | 517 | 429 |
| Chrominance AC | 1071 | 797 | 670 | 603 | 593 | 468 | 349 |

## VI. DISCUSSION

The results in Table III are much smaller than the crude upper bound $l_0 = 1642$ (5). In order to further illustrate the closeness of our result for the case $SF = 1/64$, we consider the 8x8 image block of Table IV.

Table IV: Pixel values for an example 8x8 block

| 252 | 61  | 199 | 116 | 120 | 203 | 71  | 99  |
|-----|-----|-----|-----|-----|-----|-----|-----|
| 61  | 18  | 34  | 231 | 2   | 254 | 111 | 68  |
| 199 | 34  | 229 | 165 | 192 | 247 | 250 | 53  |
| 116 | 231 | 165 | 244 | 136 | 9   | 59  | 4   |
| 120 | 2   | 192 | 136 | 233 | 252 | 27  | 59  |
| 203 | 254 | 247 | 9   | 252 | 4   | 16  | 174 |
| 71  | 111 | 250 | 59  | 27  | 16  | 247 | 11  |
| 99  | 68  | 53  | 4   | 59  | 174 | 11  | 1   |

After DCT and quantization, the configuration has 18 positions with size 8 and all other positions having size 7, which would yield 999 bits were these coefficients luminance AC and 936 bits were these chrominance AC. Our results in Table III for $SF = 1/64$ are as close as 14% to these bit-lengths and so must be even closer to the corresponding (unknown) maximum code-lengths, despite the enlargement of the configuration space $J$ to the Ball $B$.

In the derivation of the strategy, we have used a few features specific to the quantization table and the Huffman codes of the JPEG specification. Without these features some of the nice properties of the calculation will be lost, but the proposed method can be adapted and generalized to cope with any quantization table and/or Huffman codes. In particular, transformations other than the DCT may be handled as well by slightly adjusting the Ball $B$ and the AC Ball condition. Furthermore, the outlined general strategy can be adapted to DCT coefficients in MPEG 2 and 4 video compressions, both of which use quantization and VLC coding methods similar to JPEG Baseline coding. Although the coding methods of the transform coefficients are significantly different in H.264/AVC and VC-1 video compressions and in JPEG XR still image compression, some of the general ideas in this work might be useful in investigating the code-length behaviors of these compression methods. Details are left for future work.

# REFERENCES

[1] *Information Technology - Digital Compression Coding of Continuous Tone Still Images - Requirements and Guidelines*, ISO 10918-1 JPEG International Standard / ITU-T Recommendation T.81, 1992.

[2] W. B. Pennebaker and J. L. Mitchell, *JPEG Still Image Data Compression Standard*. New York: Van Nostrand Reinhold, 1993.

[3] N. Kingsbury. (2004, January 22). Image Coding. *Connexions Web*. [Online]. Available: http://cnx.org/content/col10206/1.3/

[4] S.R. Smoot, L.A. Rowe, "Study of DCT coefficient distributions" in *Proc. SPIE Symp. Electronic Imaging,* vol. 2657, San Jose, CA, 1996, pp. 403-411.

[5] T. Eude, R. Grisel, H. Cherifi, R. Debrie, "On the distribution of the DCT coefficients," in *Proc. IEEE Int. Conf. ASSP*, Adelaide, Australia, April 1994, pp. V.365-V.368.

[6] S. D. Miller, P. Smidth, C. H. Coleman, "Data compression using a feedforward quantization estimator," U.S. patent 5 146 324, Sep. 8, 1992.

[7] A. Bruna, S. Smith, F. Vella, and F. Naccari, "JPEG Rate Control Algorithm for Multimedia," in *Proc. IEEE Int. Symp. Consumer Electronics,* Reading, UK, Sep. 2004, pp. 114-117.

[8]  T. Ohta, T. Ikuma, "Electronic still camera and method operating same," US patent application publication 2001/0000969A1, May 10, 2001.

[9]  E. Hamilton, "JPEG File Interchange Format Version 1.02," C-Cube Microsystems, Sep. 1, 1992, http://www.jpeg.org/public/jfif.pdf.

[10] "Digital Still Camera Image File Format Proposal (Exif) Version 1.0 Mar. 24," JEIDA / Electronic Still Camera Working Group, 1995.

[11] G. Strang, "The Discrete Cosine Transform," *SIAM Review,* vol. 41, pp. 135-47, 1999.